# Early Output Hybrid Input Encoded Asynchronous Full Adder and Relative-Timed Ripple Carry Adder

P. Balasubramanian
School of Computer Science and Engineering
Nanyang Technological University
Singapore 639798
balasubramanian@ntu.edu.sg

K. Prasad
Department of Electrical and Electronic Engineering
Auckland University of Technology
Auckland 1142, New Zealand
krishnamachar.prasad@aut.ac.nz

*Abstract*—This paper presents a new early output hybrid input encoded asynchronous full adder designed using dual-rail and 1-of-4 delay-insensitive data codes. The proposed full adder when cascaded to form a ripple carry adder (RCA) necessitates the use of a small relative-timing assumption with respect to the internal carries, which is independent of the RCA size. The forward latency of the proposed hybrid input encoded full adder based RCA is data-dependent while its reverse latency is the least equaling the propagation delay of just one full adder. Compared to the best of the existing hybrid input encoded full adders based 32-bit RCAs, the proposed early output hybrid input encoded full adder based 32-bit RCA enables respective reductions in forward latency and area by 7.9% and 5.6% whilst dissipating the same average power; in terms of the theoretically computed cycle time, the latter reports a 10.9% reduction compared to the former.

## I. Introduction

Asynchronous circuit design using delay-insensitive data codes and a 4-phase return-to-zero (RTZ) handshake protocol is acclaimed to be a strong contender and/or a necessary supplement to mainstream synchronous circuit design by the International Technology Roadmap for Semiconductors (ITRS) design report [1]. Design for variability has been labeled as an important design challenge in the nanoscale electronics regime by the ITRS design report and in this backdrop, asynchronous circuit design based on delay-insensitive codes is attractive due to its inherent robustness to voltage, temperature and parameter variations [2] [3].

This paper presents the novel design of an early output hybrid input encoded asynchronous full adder which when cascaded to form a RCA results in less forward latency and cycle time and occupies less area compared to its existing counterparts whilst dissipating similar power. We shall first discuss some preliminaries before presenting the proposed asynchronous full adder. The dual-rail or 1-of-2 code is the simplest member of the generic family of delay-insensitive data codes [4]. In a dual-rail code, a valid data on a data wire W is represented using 2 data wires W1 and W0 as: W = 1 is represented by W1 = 1 and W0 = 0, and W = 0 is represented by W1 = 0 and W0 = 1; these two represent valid data. W1 = W0 = 0 is called the spacer, and W1 = W0 = 1 is invalid. The 1-of-4 code is used to encode two data wires (X and Y) using 4 data wires F0, F1, F2 and F3 as follows: X = Y = 0 is specified by F0 = 1; X = 0, Y = 1 is specified by F1 = 1; X = 1, Y = 0 is specified by F2 = 1 and X = Y = 1 is specified by F3 = 1. Only one of F0, F1, F2, F3 is asserted as 1 during the valid data phase. The spacer is represented by F0 to F3 all being 0s, and F0 to F3 cannot be all 1s simultaneously as it is invalid.

Strong-indication asynchronous circuits wait to receive all the input data before commencing data processing to produce the output data [5] [6]. Weak-indication asynchronous circuits tend to produce some output data after receiving even a subset of the input data but only after receiving all the input data, all the output data are produced [5] [7]. Early output asynchronous circuits could produce all the output data after receiving just a subset of the input data [8]. Early output asynchronous circuits can be further classified as early set or early reset type. If all the outputs of an early output asynchronous circuit acquire valid data after the application of just a subset of the valid inputs, it is said to be of early set type. On the other hand, if all the outputs of an early output asynchronous circuit assume the spacer state after the application of just a subset of the spacer inputs it is said to be of early reset type. The RTZ handshake protocol implies the RTZ of all the data wires (i.e. the assumption of the spacer state) after every application of valid input data [2].

## II. Proposed Full Adder – Design and Operation

Let (A0, A1), (B0, B1) and (CIN0, CIN1) denote the dual-rail full adder inputs, and (SUM0, SUM1), (COUT0, COUT1) denote the dual-rail full adder outputs. Hybrid input encoding implies the use of at least two delay-insensitive data encoding schemes, here, dual-rail and 1-of-4 codes for data encoding. The dual-rail augend and addend inputs of the full adder viz. (A0, A1) and (B0, B1) are 1-of-4 encoded, while the carry input, carry output and sum output are dual-rail encoded.

The equations governing the hybrid input encoded full adder are given by (1) to (4). Equations (1) to (4) are in disjoint sum of products/sum of disjoint products form [9] – [11], where the logical conjunction of any two product terms results in null (i.e. binary 0). Moreover, (1) to (4) satisfy the monotonic cover constraint thereby only one product term in a logical expression is activated at a time when valid input data is applied in the valid data phase.

$$SUM1 = E1CIN0 + E2CIN0 + E0CIN1 + E3CIN1 \quad (1)$$

$$SUM0 = E1CIN1 + E2CIN1 + E0CIN0 + E3CIN0 \quad (2)$$

$$COUT1 = E1CIN1 + E2CIN1 + E3 \quad (3)$$

$$COUT0 = E1CIN0 + E2CIN0 + E0 \quad (4)$$

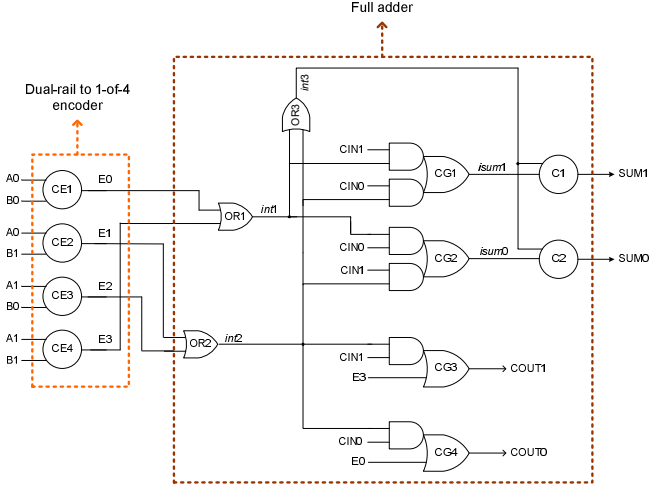

Fig. 1 Proposed hybrid input encoded early output asynchronous full adder

Fig. 1, which shows the proposed hybrid input encoded early output asynchronous full adder is the logic optimized synthesis of (1) to (4) by using simple and complex logic gates of a 32/28nm digital cell library [12]. The dual-rail to 1-of-4 encoder circuit is shown within the orange box in dotted lines, and the proposed full adder is shown within the brown box in dotted lines in Fig. 1. Logic redundancy [13] is implicit in the proposed full adder. In Fig. 1, CE1 to CE4 and C1, C2 represent 2-input C-elements. The C-element outputs 1 (0) when all its inputs are 1s (0s), and it maintains its existing steady-state otherwise. The 2-input C-element is realized using the AO222 complex gate with feedback. Gates CG1, CG2 are AO22 complex gates, and gates CG3, CG4 are AO21 complex gates. OR1, OR2 and OR3 are the simple gates. $int1$, $int2$, $int3$, $isum1$ and $isum0$ are the internal outputs. The internal outputs $isum1$ and $isum0$ are logically equivalent to the primary outputs SUM1 and SUM0 respectively. The operation of the proposed asynchronous full adder is described as follows by discussing carry-propagate, generate and kill modes with respect to Fig. 1.

*A. Carry-propagate mode*

The carry-propagate mode is specified by A0 = B1 = 1, i.e. E1 = 1 or A1 = B0 = 1, i.e. E2 = 1. If E1 or E2 is 1 during the valid data phase, OR2 will output 1 on $int2$, followed by an output of 1 on $int3$. Depending on whether CIN0 or CIN1 is 1, CG1 or CG2 is activated and an output of 1 is produced on $isum1$ or $isum0$ respectively. Subsequent to the production of 1 on $isum1$ or $isum0$, since $int3$ is also 1, the primary output SUM1 or SUM0 then evaluates to 1. Also, since $int2$ is 1, depending on whether CIN0 or CIN1 is 1, COUT0 or COUT1 evaluates to 1. In the subsequent RTZ phase, if E1 or E2, whichever was 1 earlier returns to 0 after the RTZ of A0 and B1 or A1 and B0, $int2$ would RTZ, followed by $int3$ returning to 0. Now, regardless of whether CIN0 or CIN1 whichever was 1 earlier returning to 0, $isum1$ or $isum0$, whichever was 1 earlier returns to 0. Since $int3$ and $isum1$ or $isum0$ have returned to 0, the primary output SUM1 or SUM0 whichever was 1 earlier returns to 0. Further, after $int2$ has returned to 0, COUT1 or COUT0, whichever was 1 earlier, returns to 0. Hence, irrespective of the RTZ of the carry input (CIN0/CIN1), both the sum and carry outputs of the proposed full adder could RTZ thus demonstrating its early reset nature.

*B. Carry-generate mode*

The carry-generate mode is specified by A1 = B1 = 1, i.e. E3 = 1. If E3 is 1 during the valid data phase, OR1 will output 1 on $int1$, followed by $int3$ outputting 1. Since $int1$ is 1 now, depending on whether CIN1 or CIN0 is asserted as 1, $isum1$ or $isum0$ is asserted as 1, followed by SUM1 or SUM0 being asserted as 1. Further, since E3 is 1, CG3 is activated and COUT1 is asserted as 1. In the following RTZ phase, once E3 returns to 0 after the RTZ of A1 and B1, $int1$ and subsequently $int3$ would also RTZ. Moreover, after $int1$ returns to 0, $isum1$ or $isum0$, whichever was 1 earlier would RTZ. This would be followed by SUM1 or SUM0 whichever was 1 earlier returning to 0. Also, after E3 returns to 0, COUT1 returns to 0. Hence, we find that the sum and carry outputs of the proposed full adder could RTZ regardless of the RTZ of the carry input thus demonstrating its early output viz. early reset nature.

*C. Carry-kill mode*

The carry-kill mode is specified by A0 = B0 = 1, i.e. E0 = 1. If E0 is 1 during the valid data phase, OR1 will output 1 on $int1$, followed by an output of 1 on $int3$. After $int1$ becomes 1, depending on whether CIN1 or CIN0 is asserted as 1, $isum1$ or $isum0$ is asserted as 1, followed by SUM1 or SUM0 being asserted as 1. Further, since E0 is 1, CG4 is activated and COUT0 is asserted as 1. In the following RTZ phase, once E0 returns to 0 after the RTZ of A0 and B0, $int1$ and subsequently $int3$ also RTZ. Moreover, after $int1$ returns to 0, $isum1$ or $isum0$, whichever was 1 earlier would RTZ. This would be followed by SUM1 or SUM0 whichever was 1 earlier returning to 0. Also, after E0 returns to 0, COUT0 returns to 0. Thus, the sum and carry outputs of the proposed full adder could RTZ regardless of the RTZ of the carry input once again demonstrating its early output, i.e. early reset nature.

III. RELATIVE-TIMED RCA

An important issue would arise when the proposed early output full adder is duplicated and cascaded to form an RCA, an example of which is shown in Fig. 2. Fig. 2 shows a 2-bit asynchronous RCA formed by cascading two stages of the proposed full adder. The dual-rail to 1-of-4 encoder circuits are not shown in Fig. 2 for the sake of simplicity and discussion. E0 to E3 are the 1-of-4 encoded data inputs, CIN01 and CIN00 is the dual-rail encoded carry input, and SUM01, SUM00 and COUT01 and COUT00 are the dual-rail encoded sum and carry outputs of the least significant full adder. On the other hand, E4 to E7 are the 1-of-4 encoded data inputs, COUT01 and COUT00 is the dual-rail encoded carry input, and SUM11, SUM10 and COUT11 and COUT10 are the dual-rail encoded sum and carry outputs of the most significant full adder.

In Fig. 2a, the red lines indicate a sample application of valid data inputs and the production of the corresponding valid data outputs during the valid data phase. Note that the carry-propagate mode is active in both the full adders since E1 and E5 are 1. In the least significant full adder, CIN00 is 1 and hence COUT00 becomes 1. In the most significant full adder, since COUT00 is 1, therefore COUT10 becomes 1. In Fig. 2b, the blue lines indicate an example partial RTZ of a subset of primary inputs of the 2-bit RCA and the following RTZ of all the primary outputs. It is assumed in Fig. 2b that E1 and E5 alone have returned to 0, but not the carry input CIN00 or the internal carry COUT00. It is seen that after E1 returns to 0 in the least significant full adder, OR2 outputs 0 and OR3 also outputs 0. CG1 outputs 0 and hence SUM01 returns to 0. After E5 returns to 0 in the most significant full adder, OR5 outputs 0 and OR6 also outputs 0. Since CG5 also outputs 0, SUM11 returns to 0. It is shown that CG8 outputs 0, i.e. COUT10 returns to 0. It is observed that with E1 and E5 alone returning to 0, and regardless of the RTZ of the primary carry input (CIN00) and the internal carry (COUT00), the primary sum and carry outputs of the 2-bit RCA RTZ reflecting early reset.

Given this, the late RTZ of CIN00 would not give rise to a wire orphan (i.e. unacknowledged transition on a wire) [8] [14] since the completion detector preceding the 2-bit RCA would indicate the RTZ of the primary carry input. The completion detector [2] is made up of an array of OR gates with a OR gate dedicated to combine the corresponding rails of an encoded input and the outputs of the array of such OR gates are combined using a C-element tree. The non-RTZ of the internal carry COUT00 might give rise to the problem of gate orphan [8] [14], where the gate orphan implies an unacknowledged transition on a gate output node. To overcome the likelihood of any gate orphan, a relative-timing assumption [15] is made in the region highlighted in Fig. 2b that the internal carry returns to 0 before the corresponding sum output returns to 0, i.e. the relative-timing assumption is that COUT00 returns to 0 before SUM11 returns to 0. This relative-timing assumption concerns maximum of only 2 full adder stages in any *n*-bit asynchronous RCA. So the 2-bit RCA in Fig. 2 is said to be relative-timed. However, despite the relative-timing assumption made, it is to be noted that the successive transitions within the circuit are all monotonic [16], i.e. there would be a wave of monotonically increasing (i.e. rising) transitions within the RCA during the valid data phase followed by a opposite wave of monotonically decreasing (i.e. RTZ) transitions within the RCA during the RTZ phase. Thus, imposition of the relative-timing assumption does not affect the input-output relation within the RCA.

To theoretically estimate the magnitude of relative-timing assumption to be made, we refer to the cell library information given in [12]. Note that we consider only the minimum size gates corresponding to [12] for this discussion. Referring to Fig. 2b, the critical path traversed for the direct RTZ of the sum output of a full adder consists of a 2-input OR gate viz. OR4 or OR5, an AO22 complex gate viz. CG5 or CG6, and a 2-input C-element viz. C7 or C8. The propagation delay associated with this critical path based on [12] is computed as 0.238ns. On the other hand, the critical path traversed for the indirect RTZ of the sum output of a full adder based on the carry input supplied from the preceding full adder consists of a 2-input OR gate viz. OR2, an AO21 complex gate viz. CG3 or CG4, an AO22 complex gate viz. CG5 or CG6, and a 2-input C-element viz. C7 or C8. Hence the propagation delay encountered for the indirect RTZ of the sum output of a full adder stage is estimated to be 0.301ns. Therefore, there is a small negative timing slack of 0.063ns, which is the magnitude of relative-timing assumption required for any $n$-bit asynchronous RCA constructed using the proposed full adder. The negative timing slack may however be reduced by utilizing larger size i.e. high-speed gate(s) selectively for synthesizing the carry output logic of the proposed full adder.

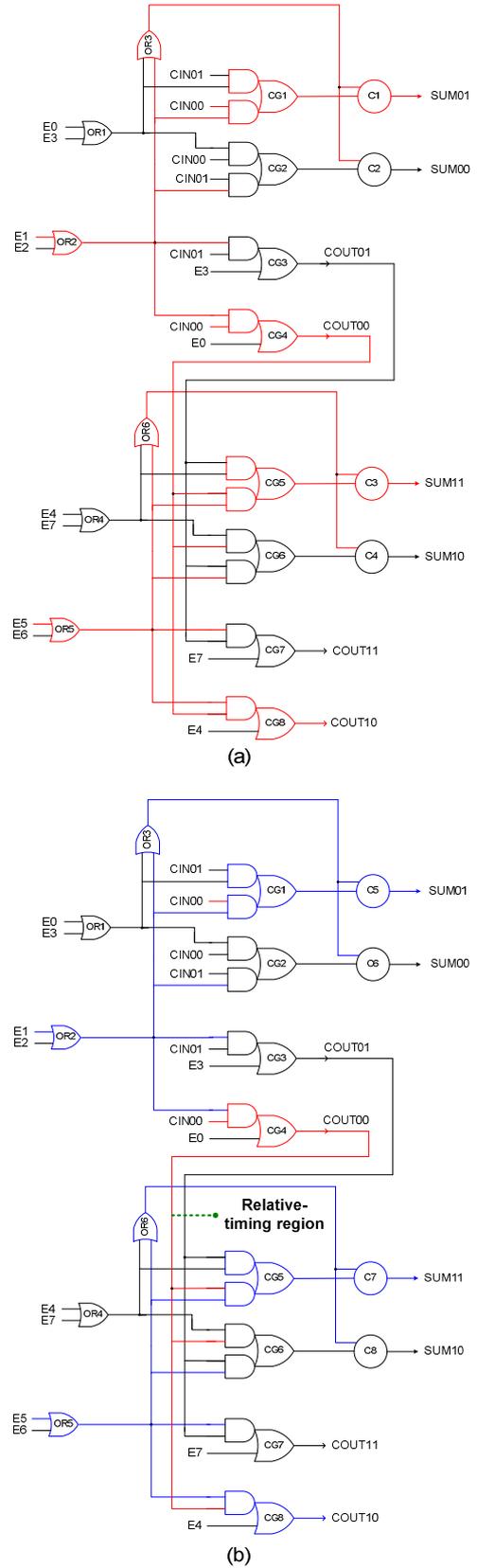

Fig. 2 A 2-bit relative-timed RCA constructed using the proposed early output hybrid input encoded asynchronous full adder

## IV. SIMULATION RESULTS AND CONCLUSION

A number of 32-bit asynchronous RCAs based on hybrid input encoded full adders corresponding to [17] – [20] and the proposed full adder was constructed using the cell library elements of a 32/28nm CMOS process [12]. Safe quasi-delay-insensitive logic decomposition [21] wherever necessary was performed to ensure gate orphan freedom. More than 1000 random input vectors were supplied to the asynchronous RCAs at time intervals of 20ns through test benches. The .vcd files generated through the simulations were subsequently used to estimate the average power dissipation. The area and forward latency (i.e. critical path delay) of the asynchronous RCAs were also estimated using Synopsys tools. To calculate the cycle times of various asynchronous RCAs, their respective forward latencies were averaged and are multiplied by the corresponding time complexities of forward and reverse latency metrics given in [22]. In a similar manner, the cycle times of different asynchronous RCAs commensurate with their actual carry propagation length(s) can be computed.

TABLE 1. (FORWARD) LATENCY, CYCLE TIME, AVERAGE POWER, AND AREA PARAMETERS OF DIFFERENT HYBRID INPUT ENCODED 32-BIT ASYNCHRONOUS RCAS. THE RCA TYPE IS MENTIONED WITHIN BRACKETS IN THE 1ST COLUMN

| RCA and its type | Latency (ns) | Cycle time (ns) | Area ($\mu m^2$) | Power ($\mu W$) |
|---|---|---|---|---|
| Reference [17] (Strong-indication) | 9.24 | 18.48 | 2504.60 | 2179 |
| Reference [17] (Weak-indication) | 8.23 | 16.46 | 2423.27 | 2175 |
| Reference [18] (Strong-indication) | 9.22 | 18.44 | 2293.14 | 2171 |
| Reference [19] (Weak-indication) | 7.21 | 14.42 | 2016.63 | 2170 |
| Reference [20] – Non-redundant logic (Weak-indication) | 7.06 | 7.50 | 2016.63 | 2170 |
| Reference [13] – Redundant logic (Weak-indication) | 3.28 | 3.49 | 2049.16 | 2170 |
| This work (Relative-timed) | 3.02 | 3.11 | 1935.30 | 2173 |

From Table 1, it is clear that the latency, cycle time and area of the proposed full adder based asynchronous RCA are the least in comparison with the optimized latency, cycle time and area metrics of an RCA constructed using the full adder of [13] which explicitly features redundant logic – thanks due to relative-timing. With respect to average power, almost all the RCAs dissipate more or less an equal value, and this is because the full adders of all the RCAs satisfy the monotonic cover constraint [2], whereby specific signal path(s) are activated between the primary inputs and primary outputs for each input vector applied. Hence in comparison with its best competitor, the proposed full adder based asynchronous RCA enables respective reductions in forward latency, cycle time and area by 7.9%, 10.9% and 5.6%.